# A New Complete Class Complexity Metric

[*1]Vinay Singh [2]Vandana Bhattacherjee
[1]*Usha Martin Academy, Booty More, Ranchi, India*
[2]*Department of CS & E, Birla Institute of Technology, Ranchi, India*
Email: [1]mailtovsingh@yahoo.co.in, [2] vbhattacharya@bitmesra.ac.in

***Abstract.*** **Software complexity metric is essential for minimizing the cost of software maintenance. Package level and system level complexity cannot be measured without class level complexity. This research addresses the class complexity metrics. This paper studies the existing class complexity metrics and proposes a new class complexity metric CCC (Complete class complexity metric), CCC metric is then analytically evaluated by Weyuker's property. An automated CCCMETRIC tool was developed for empirical sample of these metrics.**

**Keywords:** *Cyclomatic Complexity, Object-Oriented Design, Software Metrics, Weyuker's Property*

[1] Corresponding Author:
Vinay Singh
Faculty of Information Technology,
Usha Martin Academy, Ranchi, India
Email: mailtovsingh@yahoo.co.in    Tel:+919934321416

## 1. Introduction

Measurement of software complexity and fault proneness has been conducted since seventies. Cyclomatic Complexity is one of the most used metrics for static software.  McCabe introduces Cyclomatic Complexity which is considered a broad measurement of soundness of a program [13]. Qingfeng et.al. [10] is Software power (SP) which measures the software complexity by expanding the information entropy theory. Quinten et.al  [11] discussed the Refactoring of a software system to ensure its long-term maintainability through an open source system and discovered that periods of refactoring did not affect the cyclomatic complexity.

With the increased spread of Object-Oriented programming during nineties there were voices raised for the need of a metric suite that could take into consideration the complexity of Object-Oriented structure, including inheritance and polymorphism that are not present in functionally oriented software. Another important part that needs to be measured is the Object-Oriented Design phase. If one can quantify the design and thereby increase the quality of the design there is a lower probability of the software being flawed.

Software metrics measure different aspects of software complexity and therefore play an important role in analyzing and improving software quality. Measures of software complexity, for example metrics for coupling or cohesion, provide a means of quantifying its internal quality. Internal quality measures are those, which can be performed in terms of the software product itself, and it will be measureable during and after the creation of the software product.

 The recent drive towards Object-Oriented technology forces the growth of Object-Oriented software metrics [4]. The metrics suite proposed by Chidamber and Kemerer is one of the best-known Object-Oriented metrics [12]. Briand has conducted empirical studies to validate the Object-Oriented metrics for their effects upon program attributes and quality factors such as development or maintenance effort [7]. Alshayeb and Li predict that Object-Oriented metrics are effective (at least in some cases) in predicting design efforts [8]. Several other researchers have validated Object-Oriented metrics for effects of class size and with the change proneness of classes [2] [5] [18]. Li theoretically validated Chidamber and Kemerer metrics [12] using a metric evaluation framework proposed by Kitchenham et al [1] and discovered some of the deficiencies of metrics in the evaluation process and proposed a new suite of Object-Oriented metrics that overcome these deficiencies [17]. Another class complexity metric has been proposed by Balasubramanian[9] and is calculated as the sum of the number of instance variable in a class and the sum of the weighted static complexity of local methods in class. Rajnish and Bhattacherjee have studied the effect of class complexity upon development time [6] [14]. Mapping class level metrics on package level is validated through open source software and effect on quality attribute reusability, flexibility, understandability and extendibility has been studied in[16].The study on the coupling metrics and the validation through the NASA projects (KC1) and dependency of coupling on software defects has been studied in [15].

A metric may be validated mathematically using measurement theory, or empirically by collecting data. Measurement theory attempts to describe fundamental properties of all measures. Weyuker concentrated on finding desirable properties that these measures should satisfy [3]. The size measures can be useful in many ways: as input to prediction models, as a normalizing factor, as a way to express progress during development, and more. But there are other useful internal product attributes. Because practitioners and researchers think there may be a link between the structure of products and their quality, many are trying to measure the structural properties of software. Although the structural measures vary in what they measure and how they measure it, they are often called "complexity".

**Research Objectives:**
- Study the existing complexity metric and find the scope of proposed complexity metric.
- Empirically validate the complexity metrics along the proposed metric and finding that how the proposed metric can be a good predictor of measuring complexity.

The rest of the paper is organized as follows. The section 2 considers the WMC (Weighted Method per Class) metric from the Chidamber and Kemerer metric suit; CMC (Class Method Complexity) metric from Li's metric suite; and CC (Class Complexity) metric from Balasubramanian's metric suite. In section 3 we propose another complexity metric called complete class complexity metric for Object-Oriented design (CCC). Section 4 discussed the Weyuker's list of software metric evaluation criteria is presented. The result in section 5 presents the existing and proposed metrics, their validation and correlation with the data set. Concluding remarks are presented in section 6. An automated CCCMETRIC tool was developed for empirical sample of these metrics and the results have been presented.

## 2. Existing Class Complexity Metrics

**Weighted Method Per Class Metric of Chidamber and Kemerer**
Definition. Consider a class $C_i$ with methods $M_1, M_2, M_3 \ldots M_n$ that are defined in the class. Let $c_1, c_2, c_3 \ldots c_n$ be the complexity of the methods.
Then,

$$\text{WMC(Weighted Method per Class)} = \sum_{i=1}^{n} c_i$$

If all method complexity are considered to be unity, then WMC=n, the number of methods.
**Theoretical Basis**
WMC metric relates directly to complexity of an individual as defined by Bunge as the "numerosity of its composition" [10]. Thus, it can be said that the complexity of an object is the cardinality of its set of properties. In Object-Oriented terminology, the properties of an object include the instance variables and its methods. As mentioned in Chidamber and Kemerer, WMC relates directly to Bunge's definition of complexity of a thing, since methods are properties of object classes and complexity is determined by the cardinality of its set of properties. The number of methods is therefore, a measure of class definition as well as being an attribute of the class since attributes correspond to properties. They further mention that the number of instance variables has not been included in the definition of the metric since it was assumed that methods are more time consuming to design than instance variables [12].

**Viewpoints**

The number of methods and the complexity of methods involved is a predictor of how much time and effort is required to develop and maintain the class.

The large the number of methods in a class the greater the potential impact on children, since children will inherit all the methods defined in the class.

Classes with large number of methods are likely to be more application specific, limiting the possibility of reuse.

**Class Method Complexity metric of Li**

Definition: Li's CMC (Class Method Complexity) is the summation of the internal structural complexity of all local methods, regardless of whether they are visible outside the class or not(e.g. all the private, protected and public methods in class). This definition is essentially the same as the first definition of the WMC metric in [12]. However, the CMC metric's theoretical basis and viewpoints are significantly different from the WMC metric.

Theoretical Basis

The CMC metric captures the complexity of information hiding in the local methods of a class. This attribute is important for the creation of the class in an Object-Oriented design (OOD) because the complexity of the information hiding gives an indication of the amount of effort needed to design, implement, test and maintain the class.

**Viewpoints.**

The CMC metric is directly linked to the effort needed to design, implement, test and maintain a class. The more complex the class methods are, the more the effort needed to design, implement, test, and maintain the methods.

The more complex the class methods are, as measured by the internal complexity of the methods, the more the effort is needed to comprehend the realization of information hiding in a class.

**Class Complexity Metric of Balasubramanian**

Definition.

Balasubramanian's CC (Class Complexity) metric, is calculated as the sum of the number of instance variable in a class and the sum of the weighted static complexity of local methods in class [9].

To measure the static complexity Balasubramanian uses McCabe's Cyclomatic Complexity [13] where the weighted result is the number of node subtracted from the sum of the number of edges in a program flow graph and the number of connected components.

### 3. Proposed Metric

Definition: The CCC (Complete Class Complexity Metric) is proposed for measuring the complexity of class in Object-Oriented Design. To calculate the CCC metric the following nine metrics are designed at class and interface level.

Table 1. Metric Description

| *METRIC* | *NAME* | *DESCRIPTION* |
|---|---|---|
| NOMT | Number of Methods | This metric is a count the methods defined in a class. |
| AVCC | Average Cyclomatic Complexity | This metric uses McCabe's cyclomatic complexity to count the average cyclomatic complexity of methods defined in a class |
| MOA | Measure of aggregation | These metric measures the extent of the part-whole relationship, realized by using attributes. The metric is a count of the number of data declarations whose types are user-defined classes. |
| EXT | External Method calls | This metric is a count of the total number of external method calls in a class. |

| NSUP | Number of Super Class | This metric is a count of the total number of ancestor classes of a given class |
|---|---|---|
| NSUB | Number of Sub Class | This metric is a count of the total number of immediate sub classes |
| INTR | Interface Implemented | This metric is a count of the number of interfaces implemented by a class. |
| PACK | Package Imported | This metric is a count of the number of packages imported in a class. |
| NQU | Number of Queries | This metric is a count of the number of return point of all the methods in a class. |

Now the CCC metric is defined as the summation of the above metrics as follows:

$$CCC = NOMT + AVCC + MOA + EXT + NSUP + NSUB + INTR + PACK + NQU$$

Complexity Metric CCC is based upon the following assumptions.
The earlier class complexity metrics WMC defined by Chidamber-Kemerer, CMC metric by Li and CC metric by Balasubramanian have focused only on Number of methods, methods complexity, and method complexity along with the number of instance variables. It has been realized that in some cases these parameters are not sufficient to determine the complexity of the class. The focus was on each dimension of a class to measure the complexity. To measure the class complexity metric we have designed nine different metrics viz. NOMT, AVCC, MOA, EXT, NSUP, NSUB, INTR, PACK and NQU and finally get the sum of all these nine metrics to find the Complete Class Complexity Metric (CCC).

**Theoretical Basis**
In object-oriented design, classes are a combination of properties, behaviours and their relationships. The number of aggregation and generalization usually measures the relationship between classes. To measure the complexity of a class the focus should be given to all these properties. The CCC metric measures the classes at method level, attribute level and their relationships. Method complexity is measured by the number of methods, their internal complexity by average cyclomatic complexity, message sent to other methods and the number of return points. The properties in a class can be primitive type variable or a reference variable. The primitive type variable is used by the functionality of a class, the reference variable also termed as the measures of aggregation have been taken as an attribute for measuring the class complexity. The depth and height of the class should also be counted as it measures the strength of the reusability so the number of super classes, sub classes and the interface implemented has also been considered for computing class complexity.

**Viewpoints**
The CCC metric involved all the possible attribute of the class and is predictor of how much time and effort is required to design and maintain the class. The value of the CCC metric is directly linked with the understandability and testability of the class. The more the value of CCC metric are, the more the effort needed to maintain the class.
The example for illustration/calculation of existing and proposed complexity metric has been taken from [29] which is shown below.

```
package dlib;
import  java.util.*;// Package=1
public class NameDB extends NamedObject // Super class=1
{
        // constructors
        NameDB (String name) // Method=1
        {
                super(name); Number //External Method call =1
        }
        private Hashtable Names = new Hashtable();// Method of aggregation=1
```

```
        /**
         * find the name associated with a number, or return null
         */
        public Object FindName (int number)       //Method=1
        {
return( Names.get(new Integer(number)));//Queries=1;External method call =1
        }

        public Object FindName (Integer number)  //Method=1
        {
return( Names.get(number));  //Queries=1 ,External Method call =1
        }

        public Object FindNumber (String name) //Method=1
        {
                return( Names.get(name)); //Queries=1 ; External Method call =1
        }

        public void AddName (String s, int n) //Method=1
        {
                Integer i = new Integer(n);
                  Names.put(s,i); Number //External Method call=1
                  Names.put(i,s);
        }

        public void AddName (String s, Object n) //Method=1
        {
                Names.put(s,n);
                  Names.put(n,s);
        }
}
```

From the above class NameDb the following value of the design metrics are NOMT=4, AVCC =1, MOA=1, IV =0, EMC=4;NS=1; NSUB=0;NPI=1;NQ=3.
CCC = 4+1+1+0+4+1+0+3=15, WMC=4 , CMC=4+1=5 and CC =4+1+0=5

### 4. Weyuker Properties

The Weyuker properties are defined below. The notations used are as follows: P, Q and R denote classes, $P + Q$ denotes combinations of classes P and Q, µ denotes the chosen metric, µ(P) denotes the value of the metric for class P, and P$\equiv$ Q (P is equivalent to Q) means that two class designs, P and Q, provide the same functionality. The definition of combination of two classes is taken here to be the same as suggested by [3], i.e., the combination of two classes results in another class whose properties (method and instance variables) are the union of the properties of the component classes. Also, "combination" stands for Weyuker's notion of "concatenation". Let µ be the metric of classes P and Q

**Property 1:** This property states that
$$(\exists P)(\exists Q)\ (\mu(P) \neq \mu(Q))$$
It ensures that no measure rates all classes to be of same metric value.

**Property 2:** Let c be a nonnegative number. Then there are finite numbers of classes with metric c. this property ensures that there is sufficient resolution in the measurement scale to be useful.

**Property 3:** There are distinct classes P and Q such that $\mu(P) = \mu(Q)$.

**Property 4:** For object-oriented system, two classes having the same functionality could have different values, it is because of classes development are program dependent.
$$(\exists P)(\exists Q)(P = Q) \& (\mu(P) \neq \mu(Q))$$

**Property 5:** When two classes are concatenated, their metric should be greater than the metrics of each of the parts.
$$(\forall P)(\forall Q)(\mu(P) \leq \mu(P+Q) \& \mu(Q) \leq \mu(P+Q))$$

**Property 6:** This property suggests non-equivalence of interaction. If there are two classes bodies of equal metric value which, when separately concatenated to a same third class, yield program of different metric value.

For class P, Q and R
$$(\exists P)(\exists Q)(\exists R)(\mu(P) = \mu(Q) \& \mu(P+R) \neq \mu(Q+R))$$

**Property 7:** This property is not applicable for object-oriented metrics [22].

**Property 8:** It specifies that "if P is a renaming of Q; then $\mu(P) = \mu(Q)$ "

**Property 9:** It suggests that metrics of a class formed by concatenating two class bodies, at least in some cases can be greater than the sum of individual metrics such that
$$(\exists P)(\exists Q)(\mu(P) + \mu(Q) < \mu(P+Q))$$

**Table 2 Analytical Evaluation Results for CCC Metric**

| Weyuker's Property number | CCC |
|---|---|
| 1 | √ |
| 2 | √ |
| 3 | √ |
| 4 | √ |
| 5 | √ |
| 6 | √ |
| 7 | Not Applicable |
| 8 | √ |
| 9 | X |
| √ Indicates that the metric satisfies the corresponding property. X Indicates that the metric does not satisfy the corresponding property. | |

## 5. Results

**Program Flow**
The program Flow of CCCMETRICS tool is summarized in figure 1. The program flow receives Java source code as input, the source code is parsed and XML file is generated. The XML file is parsed for metric calculation and generating the Excel sheet. The required tokens are matched for the metrics WMC, CMC, CC, and the new proposed metric CCC. CCCMETRICS tool will compute the final metric value and display the result.

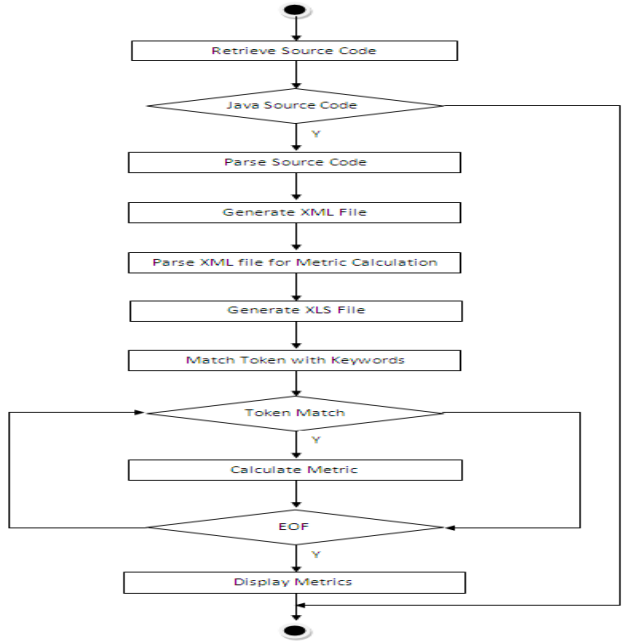

**Figure 1** Program Flow chart of CCC Metric Tool

**Evaluation and comparison of WMC, CMC, CC, CCC Metrics**

| CT | CL | Type Of Class | NM | AVCC | MOA | IV | EMC | NS | NSB | NPI | NQ | NCD | WMC | CMC | CC | CCC |
|---|---|---|---|---|---|---|---|---|---|---|---|---|---|---|---|---|
| C | JavaNames | public --- | 0 | 0 | 1 | 0 | 1 | 1 | 0 | 0 | 0 | 0 | 0 | 0.00 | 0.00 | 3.00 |
| C | NameDB | public --- | 6 | 1 | 1 | 0 | 4 | 2 | 0 | 1 | 3 | 3 | 6 | 7.00 | 7.00 | 17.00 |
| C | windowInputStream | --- --- | 4 | 1.5 | 1 | 0 | 3 | 1 | 0 | 3 | 0 | 4 | 4 | 5.50 | 5.50 | 12.00 |
| C | Deferred_PrintStream | public --- | 33 | 1.21212121212121 | 3 | 0 | 13 | 1 | 0 | 1 | 0 | 33 | 33 | 34.21 | 34.21 | 51.00 |
| C | SmallSet | public abstract | 10 | 1.4 | 0 | 1 | 4 | 1 | 0 | 0 | 3 | 7 | 10 | 11.40 | 12.40 | 18.00 |
| I | CompareFunction | public --- | 1 | 1 | 0 | 0 | 0 | 0 | 0 | 1 | 0 | 1 | 1 | 2.00 | 2.00 | 1.00 |
| I | Print_Readable | public --- | 1 | 1 | 0 | 0 | 0 | 0 | 0 | 1 | 0 | 1 | 1 | 2.00 | 2.00 | 2.00 |
| C | G | public --- | 1 | 2 | 0 | 0 | 1 | 0 | 0 | 0 | 0 | 1 | 1 | 3.00 | 3.00 | 2.00 |
| C | class_preloader | public --- | 5 | 1.8 | 2 | 2 | 6 | 0 | 0 | 1 | 0 | 5 | 5 | 6.80 | 8.80 | 14.00 |
| C | windowOutputStream | --- --- | 6 | 1.33333333333333 | 1 | 2 | 1 | 1 | 0 | 3 | 0 | 6 | 6 | 7.33 | 9.33 | 12.00 |
| C | ConsoleWindow | public --- | 23 | 3.1304347826087 | 14 | 5 | 64 | 1 | 0 | 3 | 4 | 19 | 23 | 26.13 | 31.13 | 109.00 |
| C | Readable_Printer | public --- | 12 | 2.16666666666667 | 0 | 0 | 6 | 1 | 0 | 1 | 0 | 12 | 12 | 14.17 | 14.17 | 20.00 |
| C | Deferred_PrintWriter | public --- | 32 | 1.15625 | 3 | 0 | 10 | 1 | 0 | 1 | 0 | 32 | 32 | 33.16 | 33.16 | 47.00 |
| C | NamedObject | public --- | 3 | 1 | 1 | 1 | 0 | 1 | 1 | 0 | 1 | 2 | 3 | 4.00 | 5.00 | 7.00 |
| C | KeyboardBuffer | public --- | 32 | 1.6875 | 3 | 9 | 19 | 1 | 0 | 3 | 6 | 26 | 32 | 33.69 | 42.69 | 64.00 |
| C | Queue | public --- | 5 | 1.8 | 2 | 1 | 5 | 1 | 0 | 1 | 2 | 3 | 5 | 6.80 | 7.80 | 16.00 |
| C | BaseObject | public --- | 3 | 1.33333333333333 | 0 | 0 | 0 | 0 | 7 | 0 | 3 | 0 | 3 | 4.33 | 4.33 | 13.00 |
| C | LList | public --- | 12 | 1.58333333333333 | 2 | 0 | 0 | 1 | 0 | 1 | 6 | 6 | 12 | 13.58 | 13.58 | 22.00 |

| | | |
|---|---|---|
| CT | Class Type (Class / Interface) | |
| CL | Class Name | |
| NM | Total Number of Method | |
| AVCC | Average Cyclomatic Complexity | |
| MOA | Method of Aggregation | |
| IV | Total Instance Variable | |
| EMC | External Method Call | |
| NS | Number of Super Class | |
| NSB | Number of Sub Class | |
| NPI | Number of Package Import | |
| NQ | Number of Queries | |
| NCD | Number of Command | |
| WMC | Weighted Metric Complexity | |
| CMC | Class Metric Complexity | |
| CC | Class Complexity | |
| CCC | Complete Class Complexity | |

**Figure 2 . Metric calculation sheet.**

Figure 2 is the snapshot of the parsed value of the 18 classes of DLIB system. The nine design metric value is also shown in the above sheet. The WMC, CMC, CC and CCC are calculated by the parsed value.

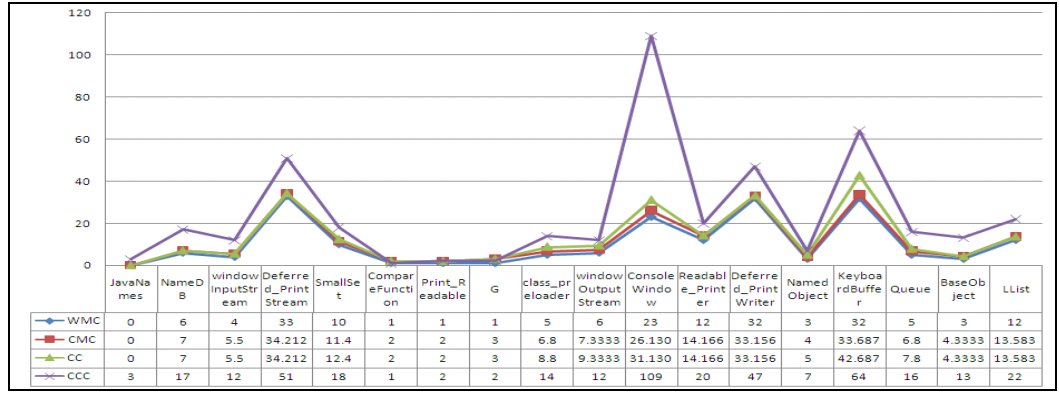

**Figure 3.Metric Comparison Chart.**

It is observed from figure 3 that CCC correlates very well with the complexity metric of WMC, CMC and CC metric. Thus, it may be used as a good complexity measure for a class design. This reaffirms the belief that complexity metric CCC is competent in measuring complexity of a class.

## 6. Conclusion and Future Scope.

In this paper, we have proposed a complete class complexity metric (CCC) based on the definition of nine metrics defined at the class and interface level. This metric has been compared with existing class complexity metrics viz. WMC, CMC and CC metric. From the illustration in section 3 and section 5, it is clear that CCC metric is able to measure the complexity of a class. The metric is evaluated through a small system (DLIB) and a comparative study proves CCC to be a better indicator of the class level complexity. A Tool was developed to calculate the CCC value and to compare it with other metrics.
The future work includes the study at system level complexity. Validation of various versions of JFree Chart is being carried on as part of our ongoing work.